\documentclass[%
reprint,
 amsmath,amssymb,
 aps,
pra,
floatfix,
]{revtex4-1}

\usepackage{graphicx}
\usepackage[caption=false]{subfig}
\usepackage{dcolumn}
\usepackage{bm}
\usepackage{hyperref}
\usepackage[mathlines]{lineno}
\usepackage{import}
\usepackage{longtable}
\usepackage[table]{xcolor}
\usepackage{array}
\newcolumntype{?}{!{\vrule width 1pt}} 

\begin{document}

\preprint{APS/123-QED}

\title{A Novel, Finite-Element Model for Spin-Exchange Optical Pumping Using  an Open-Source Code}

\author{G.M. Schrank}




\date{\today}

\begin{abstract}
A new model is presented for spin-exchange optical pumping using an open-source code, ElmerFEM-CSC. The model builds on previous models by adding the effects of alkali-vapor heterogeneity in optical pumping cells and by modeling the effects of hyperpolarized-gas wall-relaxation using a diffusion model. The code supports full, three-dimensional solutions to optical-pumping models, and solves for (1) laser absorption, (2) alkali vapor concentration, (3) fluid flow parameters, (4) thermal effects due to the pumping laser, and (5) noble gas polarization. The source code for the model is available for researchers to utilize and modify.
\end{abstract}

\pacs{Valid PACS appear here}
\maketitle


\section{\label{sec:intro}Introduction}
Spin-exchange optical pumping (SEOP) is a technique whereby the ensemble nuclear spin-angular momentum of certain noble gasses can be increased to of order 10\%. The technique is currently most notably used clinically and pre-clinically in lung imaging using MRI \cite{Oros2004HyperpolarizedMRI}, but it has also been used in the NMR characterization of porous media \cite{Terskikh2002AMaterials} and protein dynamics \cite{Schroder2013XenonAlert}. 

The physics of SEOP are described comprehensively in other places \cite{Walker2011, Appelt1998Theory129Xe}. Briefly, the technique involves two steps: (1) optical pumping of an alkali-metal vapor and (2) spin-exchange from the alkali-metal vapor to a noble-gas nuclei. In the first step, optical pumping, a beam of circularly-polarized light is directed onto a transparent cell containing a macroscopic amount of alkali metal. The cell is heated, usually to between 100-200 $^{\circ}$C, in order to vaporize some amount of alkali metal. The laser interacts with the metal vapor to create close to 100\% spin polarization of the alkali vapor.

In the second step, spin-exchange, the alkali vapor transfers spin-angular momentum to the noble gas nuclei. A gas mixture is introduced into the cell containing a noble gas and some other inert gasses, and through collisional interactions, the metal vapor transfers its spin-polarization to the noble gas. The alkali metal vapor atoms become depolarized in this interaction, but because of optical pumping, the alkali atoms are quickly repolarized.

One popular method of SEOP involves the hyperpolarization of xenon-129 ($^{129}$Xe) using rubidium (Rb) vapor. Hyperpolarized $^{129}$Xe (HP $^{129}$Xe) gas is typically produced in a continuous manner using a flow-through polarizer \cite{Driehuys1996High-volume129Xe, Ruset2006Optical129Xe, Schrank2009a}. A flow-through polarizer operates by flowing a $^{129}$Xe gas mixture through an optical pumping cell containing Rb vapor. The $^{129}$Xe spin-exchange interaction occurs on a short enough timescale that considerable $^{129}$Xe polarizations can be achieved during the short transit through the optical pumping cell.

Currently, there are no SEOP models that attempt to account for the full three-dimensional flow-dynamics of optical pumping cell geometries. Computational models for SEOP can be broken into two groups: finite difference models and finite element models (FEM). The finite difference models appear to have been the first in use, but detailed descriptions of the models do not appear extensively in the literature. Models of this type approximate flow through the cell by either a one-dimensional, plug-flow model (such as used in Ref. \cite{Freeman2014}) or two-dimensional, laminar-flow model(such as Ref. \cite{Ruset2005}). 

The first computational model of spin-exchange optical pumping to be extensively described in the literature was done so by Ref. \cite{Fink2005}. It originally described a model approximating an optical pumping cell geometry with only a half-cylinder. The model included the effects of laser heating, fluid dynamics, and heat transfer. The original model only investigated a static optical pumping cell in which the gas was not flowing. This was later extended in Ref. \cite{Fink2007} to analyze two geometries with gas flow: a half cylinder and a simplified geometry of the optical pumping cell used by Ref. \cite{Ruset2006Optical129Xe}.

Recently, Ref. \cite{Burant2018CHARACTERIZINGCONTRAST} has described an FEM fluid model of an optical pumping cell. However, it does not appear the model included SEOP, and it only included fluid flow, heat transfer, and diffusion of Rb metal vapor. 

Here, we present an open-source FEM code that attempts to model the dynamics of SEOP using the full three-dimensional geometry of a common optical-pumping-cell design. The model incorporates fluid flow, Rb vapor diffusion, thermal transfer, laser absorption, and $^{129}$Xe polarization.
\section{\label{sec:model}Model Description}
The FEM model presented here utilizes the finite-element method to numerically solve five differential equations. The author used an open-source code, ElmerFEM-CSC, to implement the model. The model simulates (1) fluid flow through the cell, (2) diffusion and transport of the Rb vapor, (3) heat transfer through the gas stream, (4) absorption of the laser by the Rb and subsequent polarization, and (5) spin-transfer between the Rb vapor and the $^{129}$Xe. Of the five modeled quantities, ElmerFEM-CSC contained modules for solving the first three. Descriptions of these three modules can be found in Ref. \cite{Raback2015}. The last two equations, for laser absorption and $^{129}$Xe polarization, were obtained by modifying existing ElmerFEM-CSC modules to accommodate the expressions that describe the dynamics of those processes. 

That this is an open-source model is a notable difference from the other models described in section \ref{sec:intro}. The other FEM models were produced using commercial codes that are not easily accessible to many researchers. Other finite-difference codes have not been formally published in an open-access environment.

The specifics of the expressions used to describe laser absorption and $^{129}$Xe polarization very closely followed Ref. \cite{Fink2005}. Major differences between the current model and other models will be highlighted in the following sections.

\subsection{\label{sec:laserasborb}Laser Absorption Model}
Laser absorption is one of the key features of an SEOP model. The expression for modeling laser absorption by the Rb-metal vapor is \cite{Wagshul1989OpticalPolarization}:
\begin{equation}
    \frac{\partial \psi(\nu,z)}{\partial z} = -n_{Rb}(z)\sigma_s(\nu)\frac{\Gamma_{SD}(z)}{\gamma_{opt}(z) + \Gamma_{SD}(z)}\psi(\nu,z)
\end{equation}
where $\psi$ is the photon flux density, $z$ is the azimuthal spacial coordinate, $\nu$ is the frequency of the light, $n_{Rb}$ is the number-density of the Rb, $\sigma_s$ is the cross-section for absorption by unpolarized Rb, $\Gamma_{SD}$ is the spin-destruction rate of the Rb, and $\gamma_{opt} = \int_0^{\infty} \psi \sigma_s \partial \nu$ is the optical pumping rate. The expression is easily solved using finite-difference methods, and it was used in many previous models. However, the presence of the integral expression is challenging for standard finite-element methods.

Instead, the FEM model presented here uses the method described by Ref. \cite{Fink2005}:
\begin{equation}
    \frac{\partial\gamma_{opt}}{\partial z} = -\beta \gamma_{opt} n_{Rb}\left(1-\frac{\gamma_{opt}}{\gamma_{opt}+\Gamma_{SD}}\right)
    \label{eq:op}
\end{equation}
where
\begin{equation}
    \beta=2\sqrt{\pi \textrm{ln}(2)}\frac{r_e f \lambda_l^2 w'\left(i\sqrt{ln(2)}(r+is)\right)}{\delta\lambda}.
    \label{eq:beta}
\end{equation}
Here, $r_e$ is the classical radius of the electron, $f$ is the oscillator strength of the Rb D-1 line, $\lambda_l$ is the laser wavelength, and $\delta\lambda_l$ is the laser-line width. $w'(Z)$ is the real part of the Faddeeva function; with $s$ denoting the ratio of the laser-to-Rb-line-frequency difference and the laser line width, and $r$ denoting the ratio of the Rb-absorption-line width and the laser-line width. 

Equation \eqref{eq:op} solves explicitly for $\gamma_{opt}$ with the assumption that the beam's spectral profile is Gaussian throughout the optical pumping cell. This is notably different from finite-difference models, in which the shape of the spectral profile changes during passage through the optical pumping cell. 

The laser absorption solver used the ElmerFEM-CSC Advection-Reaction module as its template. The equation solved by the Advection-Reaction Module is \cite{Raback2015}: 

\begin{equation}
    \frac{\partial c}{\partial t}+\vec{v}\cdot\vec{\nabla} c+\Gamma c = S
    \label{eq:advect-react}
\end{equation}
The following modifications were made to eq. \eqref{eq:advect-react}.

First, $\vec{v}$ can be constrained to be a unit vector, $\hat{n}$, pointing in the direction of the laser beam propagation. The spatial derivative of eq. \eqref{eq:op} can be rewritten as  $\hat{n} \cdot \vec{\nabla}\gamma_{opt} = \frac{\partial\gamma_{opt}}{\partial z}$.

Second, $\Gamma$ from eq. \eqref{eq:advect-react} can be set equal to:
\begin{equation}
    \Gamma=-\beta n_{Rb} \left(1-\frac{\gamma_{opt}}{\gamma_{opt}+\Gamma_{SD}}\right).
\end{equation}
The non-linear portion of the equation is solved iteratively by the Picard method \cite{Raback2015}. 

The source term, $S$, is set to 0.
\subsection{\label{sec:alkalidiff} Diffusion of the Alkali Metal}

In many previous models, the Rb-metal vapor distribution was assumed to be uniform.  In this FEM model, a diffusion model of the Rb-metal vapor is implemented using the diffusion module supplied with ElmerFEM-CSC. Simulated geometries include temperature-dependent sources and sinks, which use the Hertz-Knudsen equation for the boundary condition of the source/sink \cite{Fink2007}:
\begin{equation}
    j_{Rb}=\alpha_{Rb}\frac{p_{sat}(T)-p}{\sqrt{2\pi M_{Rb}k_B T}}
\end{equation}
where $p_{sat}(T)$ is the saturation partial pressure for a given temperature, $p$ is the instantaneous partial pressure, $M_{Rb}$ is the molecular mass of the Rb, $k_B$ is Boltzmann's constant, $T$ is the absolute temperature, and $\alpha_{Rb}$ is the evaporation coefficient of Rb. The saturation partial pressure $p_{sat}(T)$ for Rb is calculated by the Killian equation \cite{Killian1926ThermionicPotassium}. The evaporation rate $j_{Rb}$ can be positive (sources) or negative (sinks) depending on the local values of $p_{sat}(T)$ and $p$. 

Although $\alpha_{Rb}$ has not been measured for Rb, Ref. \cite{Fink2007} notes that the ideal value of $\alpha_{Rb}=1$ is expected. For all the simulations presented here, this value of $\alpha_{Rb}$ was used. For the current model, the saturation and instantaneous partial pressures were converted to the absolute mass concentration used as default in ElmerFEM-CSC diffusion module \cite{Raback2015}.

\subsection{\label{sec:wallrelax} Wall-Relaxation of HP $^{129}$Xe}
In most previous simulations, wall-relaxation is modeled as a constant term in the HP $^{129}$Xe spin-relaxation term. In this simulation, the diffusion-based model of HP $^{129}$Xe wall-relaxation is a refinement of the expression presented in Ref. \cite{Fink2005}. In that model, the wall boundary-conditions were modeled as completely depolarizing HP $^{129}$Xe spin-polarization at the walls. The authors offered an alternative model with the depolarization set to 1\% rather than 100\%. However, they noted that the lack of experimental data hindered more precise estimates.

The present model attempts to refine this approximation and connect the boundary condition at the walls to the wall-relaxation time. The wall-relaxation time (at room temperature) can easily be measured for a given cell by filling the cell with HP $^{129}$Xe and monitoring the amplitude of the HP $^{129}$Xe NMR as a function of time. It is known that this relaxation time is typically 10s of minutes \cite{Freeman2014}.

\begin{figure}[b]
    \centering
    \includegraphics[width=.45\textwidth]{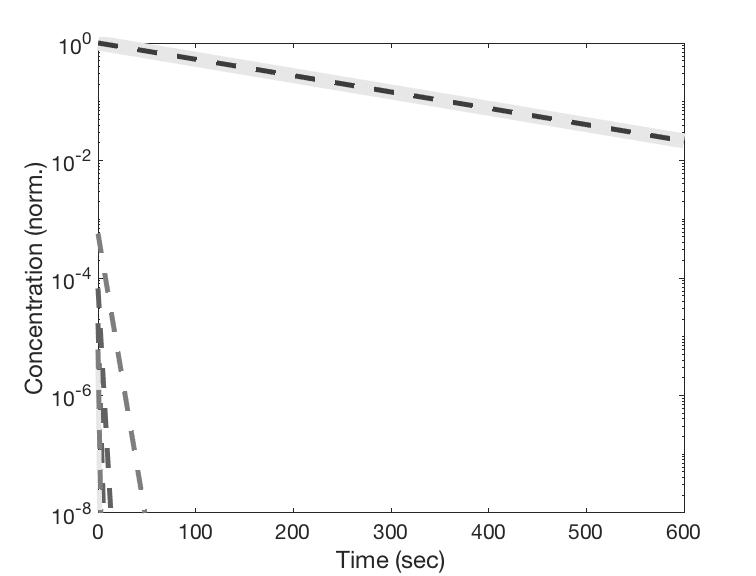}
    \caption{A plot of the decay curve (eq. \eqref{eq:poldecaycurve}). The modeled sphere had a radius of $R$=0.03 m and a diffusion coefficient of $D_{Xe}=1*10^{-5} \frac{\textrm{m}^2}{s}$. The mass transfer coefficient of $\alpha_{RW}=3.5*10^{-5}\frac{\textrm{m}}{\textrm{s}}$ was calculated using eq. \eqref{eq:masstrans} for a wall-relaxation time of $\tau_{wall}=$300 sec. The overall decay curve (solid blue) overlaps with the first-order term (dashed red). The second-order term does not contribute more than 1 part per thousand to the overall decay curve, and it decays to less than $10^{-8}$ after less than 100 seconds. The higher-order terms contribute less and decay even more quickly. This wall-relaxation time is even shorter than discussed in the text, and illustrates the robustness of the assumption at reasonable wall-relaxation times. Compare the overall decay curve with the decay curve from Figure \ref{fig:sphererelax}}.
    \label{fig:diffest}
\end{figure}

\begin{figure}
    \centering
    \includegraphics[width=.45\textwidth]{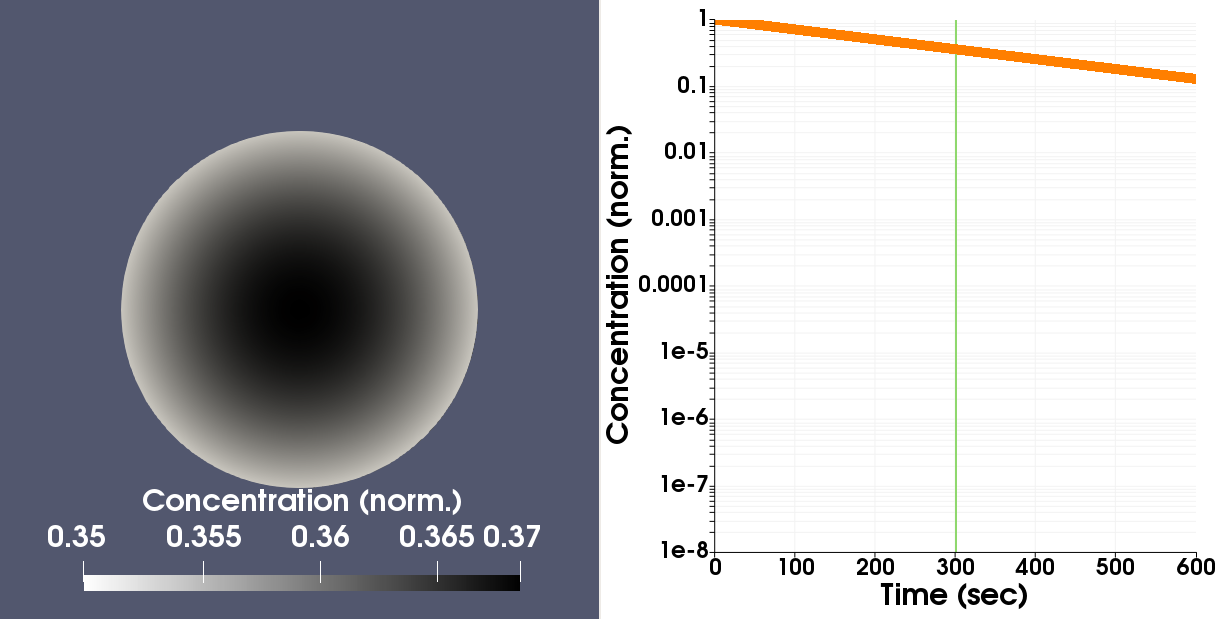}
    \caption{The visualization of an FEM diffusion model using the same parameters as those in used in Figure \ref{fig:diffest}. The left graphic depicts the concentration of the species (normalized to one) at $\tau_{wall}=300$ sec. The graph on the right is the averaged concentration as a function of time. Note that the curve is qualitatively identical to the theoretical curve in Figure \ref{fig:diffest}.}
    \label{fig:sphererelax}
\end{figure}

The present model uses the solution to the diffusion equation for surface-evaporation in a sphere. Like with the evaporation of the Rb, we wish to have an expression that relates the flux of polarization to the surface:
\begin{equation}
    j=-D_{Xe}\frac{\partial P}{\partial r}=\alpha_{RW} P,
\end{equation}
where $\alpha_{RW}$ is the so-called mass transfer coefficient.

The solution to the diffusion equation for a spherical geometry with this boundary condition is \cite{Crank1975}:
\begin{equation}
    \frac{P(r)}{P_i}=\frac{2LR}{r}\sum_{n=1}^{\infty}\frac{\textrm{exp}(-D_{Xe}\beta^2_nt/R^2)}{\beta^2_n+L(L-1)}\frac{\textrm{sin}(\beta_n r/R)}{\textrm{sin}\beta_n}
    \label{eq:diffsol}
\end{equation}
where the $\beta_n$'s are the roots of $\beta_n\textrm{cot}\left(\beta_n\right)+L-1=0$, $L=\frac{R \alpha_{RW}}{D_{Xe}}$, $P$ is the polarization after time $t$, $P_i$ is the initial polarization, and $R$ is the radius of the sphere. The total polarization, $P_{tot}$, decay curve is given by \cite{Crank1975}:
\begin{equation}
    \frac{P_{tot}}{P_i}=\sum_{n=1}^{\infty}\frac{6 L^2 \textrm{exp}(-D_{Xe}\beta^2_n t/R^2)}{\beta^2_n\left[\beta^2_n+L\left(L-1\right)\right]}.
    \label{eq:poldecaycurve}
\end{equation}

The wall-relaxation time, $\tau_{wall}$, is usually found by fitting the decaying HP$^{129}$Xe NMR amplitude to an exponential function:
\begin{equation}
    \frac{P_{tot}}{P_i}=\textrm{exp}\left(\frac{-t}{\tau_{wall}}\right).
    \label{eq:wallrelax}
\end{equation}

In the limit of $\beta_n >> \beta_1 \textrm{;} \forall n \neq 1 $, the larger $\beta_n$s can be ignored, and only the $\beta_1$ term will significantly contribute at long timescales. In this case, we can compare the time-dependent portion of just the first term of eq. \eqref{eq:poldecaycurve} to  eq. \eqref{eq:wallrelax}, and we find that $\tau_{wall}=\frac{R^2}{D_{Xe}\beta_1^2}$. 

It turns out that for typical values found in SEOP systems (i.e. $R>5$ cm, $D_{Xe}\approx0.1\frac{\textrm{cm}^2}{\textrm{s}}$ and, $\tau_{wall}>10$ min.), $\beta_1\approx1$, in which case $L<0.5$, the ratio $\frac{\beta_2}{\beta_1}\gtrapprox4$, and all other $\beta_n$s are much larger. For wall-relaxation times longer than 10 min., the ratio between the $\beta_n$s and $\beta_1$ is even greater. 

Using this approximation, we can solve for $\alpha_{RW}$ and find that:
\begin{equation}
    \alpha_{RW}=\frac{D_{Xe}}{R}\left(1-\sqrt{\frac{R^2}{D_{Xe}\tau_{wall}}}\textrm{cot}\left[\sqrt{\frac{R^2}{D_{Xe}\tau_{wall}}}\right]\right).
    \label{eq:masstrans}
\end{equation}

Equation \eqref{eq:masstrans} is a poorly behaved equation and needs to be applied with care. In particular, for a given $\tau_{wall}$, the expression has asymptotes at $R=\pi n \sqrt{D_{Xe}\tau_{wall}}$; $n=1,2,3,...$. The spacing of the asymptotes is proportional to multiples of the diffusion length. Physically, this indicates that the HP$^{129}$Xe cannot undergo wall relaxation faster than it takes for a polarized $^{129}$Xe atom at the center of the sphere to travel to the wall. Therefore, eq. \eqref{eq:masstrans} is only valid for $\tau_{wall}>\frac{R^2}{\pi^2 D_{Xe}}$.  

As an example of the validity of the assumption that the higher-order terms in eq. \eqref{eq:diffsol} and \eqref{eq:poldecaycurve} do not significantly contribute, Figure \ref{fig:diffest} shows the theoretical contributions to the decrease in concentration of the polarized species as a function of time for a sphere. The first-order term almost precisely overlays the overall decay. The higher-order terms quickly decay and do not significantly contribute to the decay curve. 

The expression from eq. \eqref{eq:masstrans} was used to model the wall-relaxation of HP$^{129}$Xe that diffused to the wall. An FEM model using this expression was checked computationally in a simple spherical model to give the correct transient relaxation time and decay curve (see Figure \ref{fig:sphererelax}). 

\subsection{\label{sec:thermtrans} Thermal Transfer through Cell Walls}
The present model attempts to capture the nuances of thermal transfer through the optical pumping cell walls. The model assumes the optical pumping cell is in a forced-air oven where the temperature of the air is held constant. Heat transfer in the walls of the cell is solved by imposing the boundary condition:
\begin{equation}
    -k\frac{\partial T}{\partial n}=\alpha_T\left(T-T_{ext}\right)
    \label{eq:tempboundary}
\end{equation}
where $k$ is the heat conductivity of the gas mixture in the cell, $T$ is the temperature at the boundary, $T_{ext}$ is the temperature at which the external flowing air is held, and $\alpha_T$ is the heat transfer coefficient. The heat transfer coefficient is calculated by combining the affects of the optical pumping wall (usually glass; thermal conductivity $k_{wall} = 1.005 \frac{\textrm{W}}{\textrm{m} \cdot\textbf{K}}$) and that of forced-air convection (heat transfer coefficient $\alpha_{air} \approx 35 \frac{\textrm{W}}{\textrm{m}^2 \cdot\textbf{K}}$). These two terms can be combined into a single heat transfer coefficient boundary condition by using the expression \cite{Bird2007}:
\begin{equation}
    \frac{1}{\alpha_T}=\frac{t_{wall}}{k_{wall}}+\frac{1}{\alpha_{air}}
    \label{eq:overheattranscoef}
\end{equation}
where $t_{wall}$ is the thickness of the wall. The expression assumes that the contact area of the different boundaries are approximately equal, which is true when $t_{wall}$ is small compared with the other linear dimensions of the optical-pumping cell.

\subsection{\label{sec:modelother} Other Considerations}
Other quantities, such as viscosity and heat capacity, were calculated using standard expressions (see Table \ref{tab:simexp}). These quantities were used as inputs for the standard modules in ElmerFEM-CSC. For more information, see appendix \ref{sec:modeldetails}.
\section{\label{sec:verification}Verification}

Although the model contains the relevant SEOP physics, it is necessary to verify that the model is providing an adequate computational solution to those expressions. In order to confirm that the FEM model conformed to accepted results, a comparison study was run with the model described in Ref. \cite{Freeman2014}, the Freeman model.

The Freeman model differs from the model described here in several important ways. The Freeman model uses a finite-difference method rather than a finite-element method. The Freeman model is a one-dimensional simulation. The fluid velocity, temperature, and the Rb-density distribution are approximated as uniform in the Freeman model. The laser spectral profile used in the Freeman model is not constrained to be Gaussian; rather the spectral profile is discretized and calculated at each spacial node in the model. Finally, in the Freeman model, the $^{129}$Xe polarization is calculated using the average Rb polarization rather than Rb polarization at each spacial node.

It was possible to reproduce aspects of this model in the infrastructure of the FEM model described here. The geometry of the FEM model was drawn as a cylinder. The Navier-Stokes fluid model, heat transfer model, and Rb-diffusion model were disabled. They were replaced with uniform axial flow, a uniform temperature, and a homogeneous Rb number-density. In addition, for this comparison, the wall-relaxation described in section \ref{sec:wallrelax} was replaced by an additional term in the HP$^{129}$Xe relaxation expression. The outlet $^{129}$Xe polarization of this FEM model was compared with the predicted $^{129}$Xe polarization from the Freeman model.  

\begin{figure}
    \centering
    \includegraphics[width=0.5\textwidth]{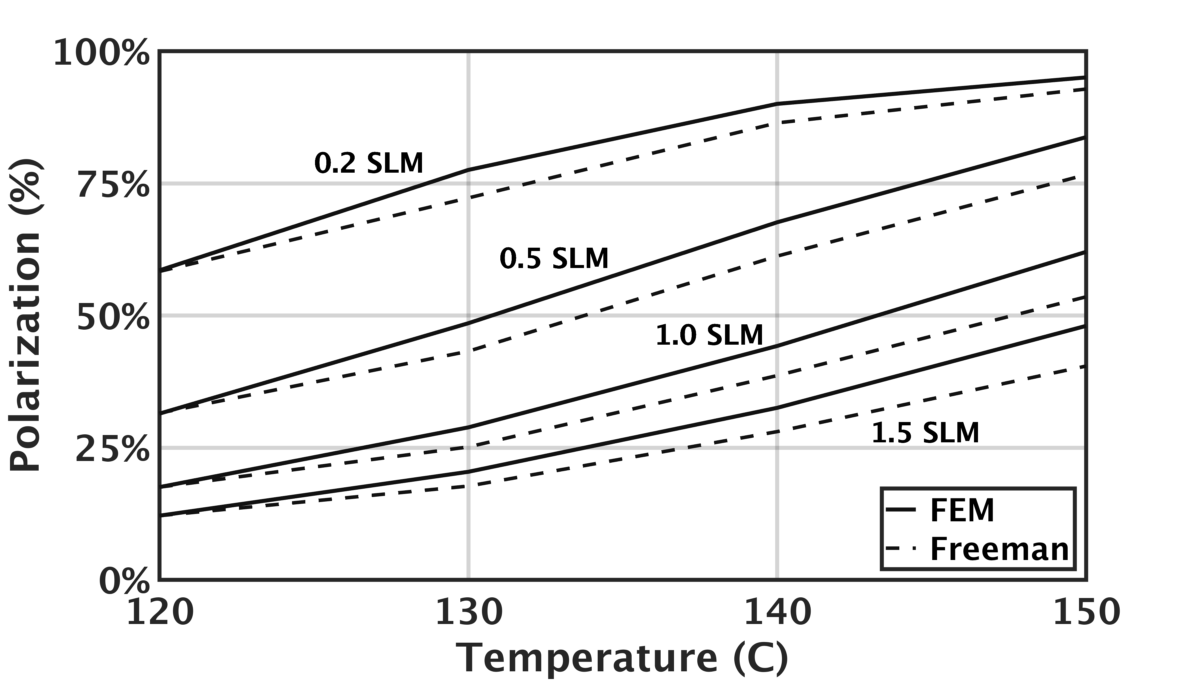}
    \caption{The predicted $^{129}$Xe polarization from the Freeman model and the FEM model as a function of temperature and flow velocity using substantially similar assumptions regarding flow, temperature distribution, and Rb number-density distribution. Each curve represents a different flow velocity (0.2, 0.5, 1.0, and 1.5 SLM) from a different model. The dotted lines are the predictions of the Freeman model, and the solid lines are the predictions of the FEM model. The Freeman model almost always predicts a lower polarization than the FEM model. This is thought to be because the FEM model includes a combination of a diffusion model for the $^{129}$Xe polarization solver and a heterogeneous Rb polarization spatial-distribution.}
    \label{fig:modelcomparison}
\end{figure}

The results of this comparison are shown in Figure \ref{fig:modelcomparison}. The two models predicted very close to the same polarization at low temperature for all flow velocities. The discrepancy between the models increases with both the temperature and the flow velocity. The Freeman model nearly always predicts a lower polarization. The largest absolute and relative discrepancies are 9 percentage-points (16\%) and 19\% (8 percentage-points), respectively.

A couple of explanations for this discrepancy were considered and rejected. One explanation for this discrepancy is that the calculated laser absorption differs in the two models. The Freeman model does not enforce the assumption of a Gaussian spectral profile of the laser, while the FEM model described here does. The assumption of a Gaussian spectral profile might influence the calculated Rb polarization, and thus the final $^{129}$Xe polarization. However, when a comparison of the 0.2 SLM flow-rate series was conducted, it was both found that the average calculated Rb polarization differed by no more than 0.2 percentage-points (0.2\%) and that the FEM model usually predicted a \textit{lower} Rb polarization. Thus, it seems unlikely that the Gaussian spectral profile, which is enforced by the FEM model, would result in a higher prediction for the $^{129}$Xe polarization.  

Another possible cause for the discrepancy between the two models was the Rb polarization spacial-distribution that existed in FEM model. Although the Freeman model does calculate a spacial-distribution for the Rb polarization, the final $^{129}$Xe polarization calculation uses a spatially-averaged value. In order to test if this was the cause of the discrepancy, the  flow direction in the FEM model was reversed so that it was directed parallel to the direction of laser propagation instead of anti-parallel. 

The Rb polarization in the FEM model is highest where the laser first enters the cell geometry. Therefore, reversing the direction of the flow in the model changes the spatial-distribution of the Rb polarization for the simulated $^{129}$Xe. In the anti-parallel flow configuration, the gas will first encounter Rb with a lower-than-average Rb polarization and progress to a region of higher-than-average Rb polarization. In the parallel flow configuration, this order is reversed.

Although the reversed-flow FEM model did result in a  lower prediction of the $^{129}$Xe polarization, this affect alone was not sufficient to account for the discrepancy between the Freeman model and the FEM model. The predicted $^{129}$Xe polarization at 150 $^{\circ}$C from the FEM model decreased by only $\sim$1 percentage point when the flow was reversed. 

A final possible cause, which was considered, was that the FEM model incorporates diffusion of the $^{129}$Xe polarization while the Freeman model does not. This difference, coupled with a Rb polarization spatial-distribution discussed above, will cause the FEM model to predict higher $^{129}$Xe polarizations than the Freeman model by significant amounts. The solution to the one-dimensional \textit{diffusion-reaction equation}, assuming a linear Rb polarization as a function of axial position, causes the final predicted $^{129}$Xe polarization to fall off linearly as a function of gas velocity. Solving the \textit{advection-reaction equation}, which is used in the Freeman model, with a Rb polarization independent of axial position causes the final predicted $^{129}$Xe polarization to fall off exponentially with gas velocity. The details of this derivation are provided in appendix \ref{sec:diffderiv}. 

This final combined effect from both the diffusion-reaction equation and the Rb-spatial distribution is likely the cause of the discrepancy between the predictions of the Freeman model and the FEM model. 

\begin{figure}[tb]
    \centering
    \includegraphics[width=.25\textwidth]{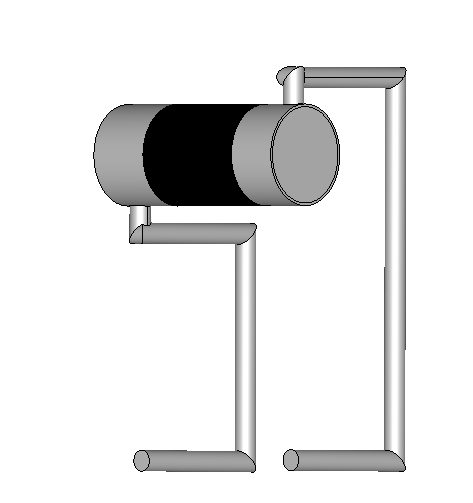}
    \caption{A visualization of the geometry chosen on which to test the FEM code. The geometry is a 100-cc standard optical pumping cell. The entire geometry (optical pumping region, inlet, and outlet) was drawn using Onshape CAD software, and a mesh was created from this geometry using Salome 9.2.1. The Rb source was modelled as a thin, cylindrical film that encircled the middle of the optical pumping region of the cell (black section in figure). An additional Rb sink was prescribed in the entire outlet tube for the geometry (not pictured).}
    \label{fig:geometry}
\end{figure}
\section{Test on a Standard SEOP Geometry\label{sec:3dtests}}

\begin{figure}[bt]
    \centering
    \includegraphics[width=0.45\textwidth]{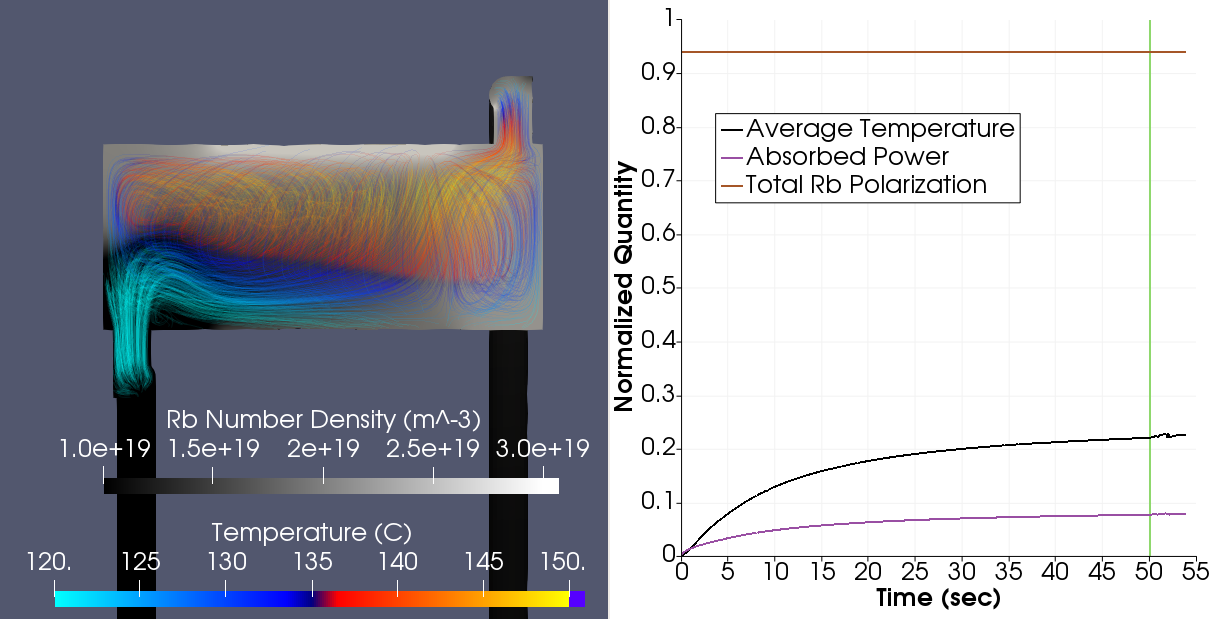}
    \caption{A visualization of the 100-cc cell simulation. The image on the left shows the predicted flow-lines for the gas as it moves through the cell. The flow lines are color-coded by the predicted temperature of the gas at each point. The slice of the cell is color-coded by the predicted Rb number-density in the cell. The graph on the right shows three of the simulated quantities averaged over the optical-pumping region of the cell: (1) Rb polarization, (2) absorbed laser power, and (3) temperature. The Rb polarization and absorbed laser power are both normalized to one. The temperature is normalized to the fraction above the set-point temperature, in this case, 110 $^{\circ}$C.}
    \label{fig:110ccfulltrans}
\end{figure}

After verification, limited tests were conducted on a standard, SEOP-cell geometry to assure that all modules would work in concert. A 100-cc-cell geometry with Rb in the optical-pumping body was chosen as the first geometry to test (see Figure \ref{fig:geometry}). The test was conducted with the following boundary conditions:
\begin{itemize}
    \itemsep 0 em
    \item Laser Power: 75 W
    \item Laser Spectral Width: 0.3 nm
    \item Laser Beam Radius: 1 mm smaller than optical pumping region radius
    \item Wall Temperature: 110 $^{\circ}$C
    \item Flow Rate: 1.5 SLM
    \item Wall-Relaxation Time: 56 min.
    \item Fraction $^{129}$Xe: 1\%
    \item Fraction N$_2$: 10\%
    \item Fraction He: 89\%
    \item Cell Pressure: 73 psig (assumed to be at sea level)
\end{itemize}

Initially, a steady-state solution for the model was attempted. However, when this failed to converge, the model was solved as a transient problem with 0.1 sec. per time-step. A total of 611 time-steps were simulated. However, due to a file-corruption error, the last 6 time-steps were discarded.  

Due to possible instability with the module, the $^{129}$Xe polarization module was not used during the transient simulation. Instead, after the transient simulation reached steady-state in (1) temperature, (2) Rb polarization, and (3) laser power absorbed, the last transient solution was used as the initial conditions for a steady-state calculation with the HP$^{129}$Xe polarization module enabled. Steady-state was defined as less than a 0.05\% change in all of the three other metrics.

The transient simulation shows convergence to a steady-state solution after the 611 time-steps (Figure \ref{fig:110ccfulltrans}). The predicted $^{129}$Xe polarization is comparable to the observed $^{129}$Xe polarization, made in Ref. \cite{Freeman2014}, from a 100-cc cell similar to the geometry simulated in the FEM model. In the Ref. \cite{Freeman2014} study, a 100-cc cell was observed to absorb 30\% of the incident light and produce HP$^{129}$Xe polarized to $\sim$15\% at 90 $^{\circ}$C. The FEM model predicted that  the simulated 100-cc cell would absorb 7.9\% of the incident light and produce HP$^{129}$Xe polarized to $\sim$4.7\% at 110 $^{\circ}$C. The HP$^{129}$Xe polarization for the FEM simulation was taken as the average polarization across a slice 5 cm from the edge of the outlet of the model's geometry.

The discrepancy between the observed light absorption of the 100-cc optical pumping cell and the simulated cell may be from a couple sources. First, as previously mentioned, the FEM model enforces a Gaussian spectral profile throughout the optical pumping cell. This assumption may affect the calculated total absorption of the light. The affects of the imposed Gaussian distribution are not entirely understood, however, as stated in section \ref{sec:verification}, the Gaussian spectral distribution was found to not significantly alter the average Rb polarization in the FEM model when compared with a model that does not enforce a Gaussian spectral distribution. 

Second, the Rb-source distribution in the model may not accurately describe the distribution in the optical pumping cell tested in Ref. \cite{Freeman2014}. The authors do not describe the Rb-metal distribution in their 100-cc cell (e.g. a Rb droplet, a thin layer of Rb on the side of the cell, etc.). The distribution and location of the Rb-metal in the cell may affect the details of the dynamics in the optical-pumping region.

The discrepancy between the observed HP$^{129}$Xe polarization may be similarly explained by the Rb-source distribution in the model. Because the Rb-source distribution may not reflect the conditions in the actual cell that was tested, the Rb number-density calculated in the model may be lower than the number-density that was realized in the actual cell. This would result in a lower spin-exchange rate than what was realized in the actual cell, and thus, a lower predicted HP$^{129}$Xe polarization.

\begin{figure}
    \centering
    \includegraphics[width=0.45\textwidth]{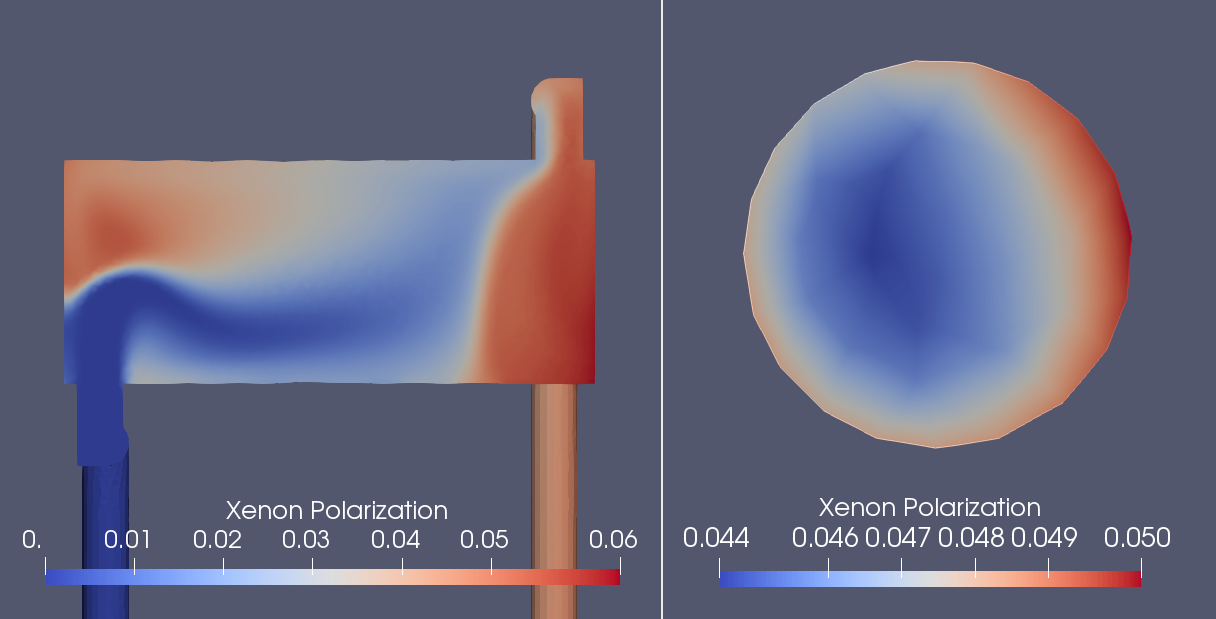}
    \caption{A visualization of the steady state solution to the HP$^{129}$Xe polarization. The visualization on the left shows the polarization in the optical pumping section of the cell. The maximum polarization is $\sim$6\%. The visualization on the right is the polarization 5 cm before the edge of the outlet of the cell. The total polarization averaged over this section is 4.7\%.}
    \label{fig:110xepol}
\end{figure}
\section{Summary and Future Work\label{sec:conclusion}}
A new open-source, FEM-based SEOP model has been coded, tested against an existing model, and tested with a three-dimensional geometry. The model provides predictions of HP$^{129}$Xe polarizations that are comparable to existing, accepted models when the conditions are limited to the scope of those existing models. The three-dimensional model predicts HP$^{129}$Xe polarizations that are comparable to existing observations. The new FEM model provides the ability to visualize important SEOP phenomena such as laser heating, Rb-vapor distribution, and gas flow. The model can compute solutions for complicated geometries including current designs for optical pumping cells.

The most immediate improvement that can be made is to add the ability to multi-thread computations. The lack of multi-threading limits the speed at which solutions can be computed, which in turn, limits transient studies to simulate only the first several hundred seconds after initialization. The three-dimensional simulation described in section \ref{sec:3dtests} required $\sim$22,000 cpu-min., and it only simulated the first $\sim$60 sec. of the cell after initialization. Important, long-term behavior cannot be thoroughly investigated. Multi-threading capabilities with appropriate increases in other computational resources could potentially decrease the amount of time required to calculate a single time-step and open up the possibility of exploring the long-term behavior of these systems.

Better approximations to spin-exchange parameters and other model parameters may increase the fidelity of the model's predictions. However, it is suspected that a major source of error in the model is likely due to the choice of Rb source distribution. Because all the parameters are correct to an order of magnitude, it is unlikely substantial gains will be realized by pursuing more accurate approximations to these parameters unless it can be shown that the current estimates of the parameters are incorrect by large margins.

Further simulations on various optical pumping cell geometries will be reported in a follow-up report.
\begin{acknowledgments}
The author gratefully acknowledges J. Cook for useful conversations regarding the coding of the simulations. Also, the author is grateful for insightful discussions with and feedback on initial drafts from B.J. Anderson, T. Barthel, B. Driehuys, and B. Saam. Finally, much of the computation time on Amazon's AWS Cloud Computing Service was supported by backers of the project from Experiments.com. The author gratefully acknowledges the financial support of all the backers of the project.
\end{acknowledgments}

\appendix
\section{Model Details\label{sec:modeldetails}}
\begin{table*}[p]
\caption{The table lists all of the expressions that are used to calculate various parameters in the model. The first column lists the term in the model. The second column lists in which model the term is specifically used. The third column lists the expression used. Note that because the model is coded in SI units, the actual implementation of some of the expression may have been multiplied by a constant to make all the units consistent. The final column lists the reference from which each expression is derived. In all cases, the notation used by the reference has been kept. For the meaning of the particular notation in the expressions, the reader should refer to the particular reference. *Note: A previous version of this pre-print incorrectly listed the expressions used for specific heat ratio and viscosity.\label{tab:simexp}}
\begin{tabular}{|l|c|c|c|}
\hline
\multicolumn{1}{|c|}{\textbf{Term}}& \textbf{Model} & \textbf{Expressions} & \textbf{Reference} \\
\hline
\begin{tabular}{@{}l@{}}Gas Mixture\\ Density\end{tabular}& 
\begin{tabular}{@{}c@{}} Navier-Stokes\\ Heat \end{tabular} &
\begin{tabular}{@{}c@{}} $\rho=\frac{P}{TR}$ \\ $R=C_p\frac{\gamma-1}{\gamma}$ \end{tabular} &
\cite{Raback2015}\\ 
\hline 
\begin{tabular}{@{}l@{}}Gas Mixture\\ Viscosity\end{tabular}& 
Navier-Stokes&
\begin{tabular}{@{}c@{}}
    $\mu = \mu'\left(\frac{T}{T'}\right)^{\frac{3}{2}}\frac{T'+S}{T+S}$ \\
    $\mu_{tot}=\sum_i x_i (\mu_i)^{\frac{1}{3}}$  
\end{tabular}&
\\
\hline 
\begin{tabular}{@{}l@{}}Heat Capacity\\ at Constant Pressure\end{tabular}&
\begin{tabular}{@{}c@{}} Navier-Stokes \\Heat \\Rb Diffusion  \end{tabular}&
$C_{p,mix} = \sum_i x_i C_{p,i}$&
\cite{NIST69}\\
\hline 
\begin{tabular}{@{}l@{}}Specific Heat\\Ratio\end{tabular}&
\begin{tabular}{@{}c@{}}Navier-Stoke \\ Heat\\ Rb Diffusion\end{tabular}& 
$\gamma = \sum_i x_i \gamma_i$&
\\
\hline 
\begin{tabular}{@{}l@{}}Alkali Evaporation\\Rate\end{tabular}&
Rb Diffusion&
$j_{Rb}=\alpha\frac{ p_{sat}-p}{\sqrt{2\pi M_{Rb} k_BT}}$&
\cite{Fink2007}\\ 
\hline 
Diffusion Constant&
\begin{tabular}{@{}c@{}}Rb Diffusion\\$^{129}$Xe Polarization\end{tabular}&
$D_{1,2}=\frac{1.8583\times10^{-7} T^{3/2}\sqrt{\frac{1}{M_1}+\frac{1}{M_2}}}{p\sigma^2_{1,2}\Omega}$&
\cite{Bird2007}\\ 
\hline
\begin{tabular}{@{}l@{}}Optical Pumping\\ Rate\end{tabular}&
Laser Absorption&
$\frac{\partial \gamma_p}{\partial z}=-\beta\gamma_p n_{Rb} \left(1-\frac{\gamma_p}{\gamma_p+\Gamma_{SD}}\right)$&
\cite{Fink2005}\\
\hline
\begin{tabular}{@{}l@{}}Alkali Spin\\ Destruction Rate\end{tabular}&
Laser Absorption&
\begin{tabular}{@{}c@{}}$\Gamma_{SD}=\Gamma_{Rb}+\Gamma_{Xe}+\Gamma_{N_2}+\Gamma_{He}+\Gamma_{VW}$\\$\Gamma_{Rb}=\kappa_{Rb}[Rb]$\\$\Gamma_{N_2}=170\left(1+\frac{T-90 ^{\circ} C}{194.36 ^{\circ} C}\right)[N_2]$\\$\Gamma_{He}=24.6\left(1+\frac{T-90 ^{\circ} C}{96.4 ^{\circ} C}\right)[He]$\\$\Gamma_{Xe}=\kappa_{Xe}[Xe]$\\$\Gamma_{VW}=\frac{6469}{f_{Xe}+1.1f_{N_2}+3.2f_{He}}$\end{tabular}&
\begin{tabular}{@{}c@{}}\cite{Ruset2005}\\ \cite{Nelson2001} \\ \cite{Walter2002} \\ \cite{Baranga1998}\end{tabular}\\
\hline
\begin{tabular}{@{}l@{}}Xenon Spin-\\Exchange Rate\end{tabular}&
$^{129}$Xe Polarization&
$\gamma_{SE}=\kappa_{SE}[Rb]$&
\cite{Fink2005}\\
\hline
\begin{tabular}{@{}l@{}}Xenon Spin-\\Relaxation Rate\end{tabular}&
$^{129}$Xe Polarization&
\begin{tabular}{@{}c@{}}$\Gamma_{SR}=\Gamma_B+\Gamma_{VW}$\\ $\Gamma_b=\kappa_{Xe}[Xe]$\\$\Gamma_{vdW}=\frac{\Gamma^{Xe}_{vdW}}{1+\frac{r[B]}{[Xe]}}$\end{tabular}&\cite{Chann2002}\\
\hline
Wall-Relaxation&
$^{129}$Xe Polarization&
$\alpha=\frac{D_{Xe}}{R}\left(1-\sqrt{\frac{R^2}{D_{Xe}\tau}}\textrm{cot}\left[\sqrt{\frac{R^2}{D_{Xe}\tau}}\right]\right)$&
\cite{Crank1975}\\
\hline
Thermal Conductivity&
Heat&
\begin{tabular}{@{}c@{}}$k=1.9881 \times 10^{-4}\frac{\sqrt{T/M}}{\sigma^2\Omega}$\\$k_{mix}=\sum_\alpha \frac{x_{\alpha}k_{\alpha}}{\sum_\beta x_\beta \phi_\beta}$\end{tabular}&
\cite{Bird2007}\\
\hline
\begin{tabular}{@{}l@{}}Heat Transfer\\ Coefficient\end{tabular}&
Heat&
$\frac{1}{U}=\frac{1}{h_0}+\sum_{j=1}^n\frac{x_j-x_{j-1}}{k_{j-1,j}}-\frac{1}{h_n}$&
\cite{Bird2007}\\
\hline
Laser Heating&
Heat&
$Q=h\nu_l n_{Rb} \gamma_p\frac{\Gamma_{SD}}{\gamma_p+\Gamma_{SD}}$&\cite{Fink2005}\\ \hline
\end{tabular}
\end{table*}
Section \ref{sec:model} was meant to highlight the important differences between the FEM model and previous models. However, readers that are interested in using the model may be interested in some more details of that model. The most recent version of the model can be found at \url{https://github.com/drschrank/elmerfem} and is free for use under the GNU license requirement found in the source code.

Although the model can be used in a steady-state formulation of the differential equations, transient simulations are frequently useful because initial simulations on some optical pumping geometries fail to converge when solving the steady-state equations. During testing, the transient time-step for the simulation was 0.1 sec/step.

Two different linear solvers with different convergence limits were used for the modules. The generalized minimal residual method (GMRES) and biconjugate gradient stabilized method (BiCGSTAB) were both used for the Navier-Stokes equation. The BiCGSTAB was used exclusively for all the other modules. Non-linear and linear convergence limits were all less than $10^{-4}$ and were usually $10^{-6}$. 

As described in section \ref{sec:model}, the model consists of five modules which are calculated in sequence. The first module which is calculated is the solution to the Navier-Stokes equation. It was provided with ElmerFEM-CSC and was used without modification. ElmerFEM-CSC provides two applicable compressiblity models for gas flow: the ``Perfect Gas'' compressibility model, which solves the full Navier-Stokes equation, and  the ``Incompressible'' compressibility model, which solves the Navier-Stokes equation with the assumption that $\vec{\nabla} \cdot j=0$. The ``Incompressible''compressiblity model requires less computational resources, and it is appropriate when the Mach number of the fluid flow is low, which is the case for all reasonable SEOP models. When using the ``Incompressible'' model, the density can be calculated as a function of temperature at each time-step. The Navier-Stokes calculation is dependent on the solution to the heat equation, which is discussed later. 

The second module which is calculated is Rb diffusion. This module uses the Advection-Diffusion module. This module was also provided with ElmerFEM-CSC and was used without modification. This module's solution is only dependent on the resulting gas flow calculated by the Navier-Stokes equation.

The third module is the laser absorption module. This module was based on the Advection-Reaction (eq. \eqref{eq:advect-react}) module, which was provided by ElmerFEM-CSC. The module was modified to use the Picard method to approximate the non-linear solution by a series of iterative steps (see Ref. \cite{Raback2015} section on linearization of the Navier-Stokes equation for an example). The specific equation solved using this method is described in section \ref{sec:laserasborb}. The solution to this module is only dependent on the solution to the previous module, the Rb diffusion module.

The time-dependent portion of eq. \eqref{eq:advect-react} is ignored by ElmerFEM-CSC solvers when the model is used to solve steady-state problems. However, for the transient simulations, the time-dependent term had to be included. To handle this, the time-dependent term was multiplied by a coefficient that is much smaller than the characteristic time-steps used in the model. This modification effectively causes each time-step to be a steady-state solution of eq. \eqref{eq:op}.

The fourth module is the $^{129}$Xe polarization module. This module is based on the Advection-Diffusion module which was provided by ElmerFEM-CSC. The method by which the equation was solved was not changed. The only changes made were the assignments to the various constant parameters of the Advection-Diffusion equation, and the solver was forced to always use the ``absolute mass'' setting because the equation in this form is easily adapted to eq. \eqref{eq:xediff}. 

The solution to this module is only dependent on the laser absorption module, and the solution does not effect any of the other modules. This allows for the possibility of running simulations with only the other four modules active and then using a final, steady-state solution of the other four modules to calculate the solution to the $^{129}$Xe polarization.

The final module is the heat equation module. This module was provided by ElmerFEM-CSC without any modification. The solution is dependent on both the laser absorption solution and the Navier-Stokes solution.

Parameters for the various modules are listed in table \ref{tab:simexp}. The values of particular constants used in the equations in the table can be found in the references listed in the table.
\section{Details of the Derivation of the Difference Between the Freeman Model and the FEM Model\label{sec:diffderiv}}
In section \ref{sec:verification}, it is stated without proof that the assumptions of the Freeman model give rise to an exponential dependence on flow rate when a uniform Rb polarization is assumed, while the assumptions of the FEM model give rise to a linear dependence on flow rate when a Rb polarization gradient is assumed. In this appendix, the derivation of that result will be given.

The FEM model uses the advection-diffusion equation to model $^{129}$Xe polarization. The form of that equation in one dimension is:
\begin{multline}
    D_{Xe}\frac{\partial^2 P_{Xe}(x)}{\partial x^2}+v\frac{\partial P_{Xe}(x)}{\partial x}+\left(\sigma+\Gamma\right)P_{Xe}(x)= \sigma P_{Rb}(x).
    \label{eq:xediff}
\end{multline}

The Rb polarization is assumed to have a linear gradient given by:
\begin{equation}
   P_{Rb}(x)=\frac{P_L-P_0}{L}x+P_0
   \label{eq:linrbdef}
\end{equation}
where $P_L$ is the highest polarization of the Rb at the point $x=L$, and $P_0$ is the lowest Rb polarization at the point $x=0$. From Ref. \cite{Walker2011}, it is clear in the limit as $L \to \infty$ that $P_{Xe} \to \frac{\sigma}{\sigma+\Gamma}P_L$. The solution to eq. \eqref{eq:xediff} is given by:
\begin{multline}
     P_{Xe}(x) =  K_1e^{x\frac{v+\sqrt{v^2-4D_{Xe}(\sigma+\Gamma)}}{2D_{Xe}}}+K_2e^{x\frac{v-\sqrt{v^2-4D_{Xe}(\sigma+\Gamma)}}{2D_{Xe}}} -\\ \frac{v \sigma (P_L-P_0)}{L(\sigma+\Gamma)^2}+\frac{\sigma(P_L-P_0)x}{L(\sigma+\Gamma)}+\frac{\sigma P_0}{(\sigma+\Gamma)}.   
\end{multline}

For $v^2>>4D_{Xe}(\sigma+\Gamma)$, this can be simplified to:
\begin{multline}
         P_{Xe}(x) =  K_1e^{-\frac{vx}{D_{Xe}}}+K_2+\\ \frac{\sigma}{\sigma+\Gamma}\left(P_{Rb}(x)-\frac{v(P_L-P_0}{L(\sigma+\Gamma)}\right).
\end{multline}

If $\frac{vL}{D_{Xe}}>>1$, then 
\begin{multline}
    P_{Xe}(L) = K_2+\frac{\sigma}{\sigma+\Gamma}\left(P_L-\frac{v(P_L-P_0}{L(\sigma+\Gamma)}\right).
\end{multline}

For $L \to \infty$, we get:
\begin{equation}
    P_{Xe}(L\to\infty) \to K_2+\frac{\sigma}{\sigma+\Gamma}P_L = \frac{\sigma}{\sigma+\Gamma}P_L,
\end{equation}
which implies:
\begin{equation}
    K_2=0.
\end{equation}

Therefore, solving the one-dimensional advection-diffusion equation for the polarization of $^{129}$Xe  at $x=L$ assuming a linear Rb polarization gradient gives:
\begin{equation}
    P_{Xe}(L) = \frac{\sigma}{\sigma+\Gamma}\left(P_L-\frac{v(P_L-P_0}{L(\sigma+\Gamma)}\right),
\end{equation}
which linearly decreases as the flow rate, $v$, increases.

The Freeman model uses eq. \eqref{eq:xediff} with the diffusion term absent and an average Rb polarization, $\bar P_{ave} = \frac{P_L+P_0}{2}$. The solution to this equation can be trivially shown to be:
\begin{equation}
    P_{Xe}(L) = \frac{\bar P_{ave} \sigma}{\sigma+\Gamma}\left(1-e^{-\frac{(\sigma+\Gamma)L}{v}}\right),
\end{equation}
which exponentially decreases as a function of $v$.

\bibliographystyle{apsrev4-1}
\bibliography{references}

\end{document}